\documentclass{article}
\usepackage{graphicx} 
\usepackage{booktabs} 
\usepackage{hyperref}
\usepackage{amsmath} 
\usepackage[nameinlink,noabbrev]{cleveref}
\usepackage{array}
\usepackage{multirow}
\usepackage{makecell}
\usepackage{subcaption}
\usepackage{floatrow}
\newfloatcommand{capbtabbox}{table}[][\FBwidth]
\usepackage[a4paper,top=2cm,bottom=3cm,left=3cm,right=3cm]{geometry}
\usepackage{amsmath}

\begin{document}
\title{\textbf{The Political Resource Curse Redux}}
\author{Hanyuan Jiang}
\date{} 
\maketitle

In \textit{the study} of the Political Resource Curse (Brollo et al.,2013), the authors identified a new channel to investigate whether the windfalls of resources are unambiguously beneficial to society, both with theory and empirical evidence. This paper revisits the framework with a new dataset. Specifically, we implemented a regression discontinuity design and difference-in-difference specification.\\

\section{Introduction}
Unlike the commonly accepted “Dutch disease” hypothesis, which reveals how a windfall of resources can reduce income via the market mechanism, or the literature about the conflict between different interest groups, the focus of our study is adverse effects on the functioning of political institutions:how such windfall exacerbates the political agency problem and deteriorates the quality of political candidates.\\

According to the extended political agency model introduced in "The Political Resource Curse"(\textit{the study}), an incumbent competes for reelection against a set of heterogeneous challengers.  When faced with an increase in non-tax government revenues (windfall), the incumbent faces a trade-off between grabbing benefits for himself versus pleasing the voters to increase the probability of reelection. One result is the moral hazard, as the electoral punishment of corruption diminishes with budget size, and this induces more frequent misbehaving (corruption). Another result is the selection effect, which will cause a decline in the average ability of the pool of political candidates. This is a by-product of the first result and of the assumption that these benefits are more valuable for political candidates with lower ability.\\

To test this theory, we use data from a sample of Brazilian municipalities. One unique aspect of the Brazilian federal transfer system, known as the Fundo de Participação dos Municipios (FPM), is that transfers to municipalities change exogenously and discontinuously at given population thresholds, with all municipalities in the same state and a given population bracket receiving the same transfers. The key empirical evidence is that municipalities just above and below these thresholds are comparable in all aspects except for the amount of transfers they receive. Under this setting, we implement a fuzzy RD design with population discontinuities as the instrument to study the effects of a discrete change in transfers between these municipalities. To complement the RD design, we also conduct a difference-in-difference estimation.\\
\\
\section{Empirical Framework}
\subsection{Federal Transfers to Municipal Governments}

According to \textit{the study}, FPM is the single most important source of municipal revenues (40\%), consisting of automatic federal transfers established by the Federal Constitution of Brazil. Our study focuses on FPM, both for its importance and its allocation depending on a discontinuous population size which is crucial for identification strategy.\\

\begin{table}[ht]
\centering
\caption{FPM Coefficients}
\begin{tabular}{lccr}
 & & &\\ 
\toprule
Population interval & FPM coefficient & Actual Transfers & Theoretical 
Transfers \\
\midrule
Below 10,189 & 0.6 & 19.68 & 18.15 \\
10,189--13,584 & 0.8 & 25.19 & 23.90 \\
13,585--16,980 & 1 & 31.25 & 30.51 \\
16,981--23,772 & 1.2 & 37.22 & 36.86 \\
23,773--30,564 & 1.4 & 43.58 & 42.86 \\
30,565--37,356 & 1.6 & 49.69 & 49.32 \\
37,357--44,148 & 1.8 & 55.68 & 55.24 \\
44,149--50,940 & 2 & 62.61 & 63.45 \\
\midrule
Total & & 33.14 & 32.27 \\
\bottomrule
\end{tabular}\\
\label{tab:Table1}
\endcenter
\textit{Source: The Political Resource Curse (Brollo et al., 2013)}
\end{table}

According to the FPM allocation mechanism, municipalities are divided into population brackets that determine the coefficients used to share total state resources earmarked for the FPM, with smaller population brackets corresponding to lower coefficients. Each state receives a different share of the total resources earmarked for FPM, two municipalities in the same population bracket receive identical transfers only if they are located in the same state. \\

\begin{figure}[t]
    \centering
    \includegraphics[width=0.5\linewidth]{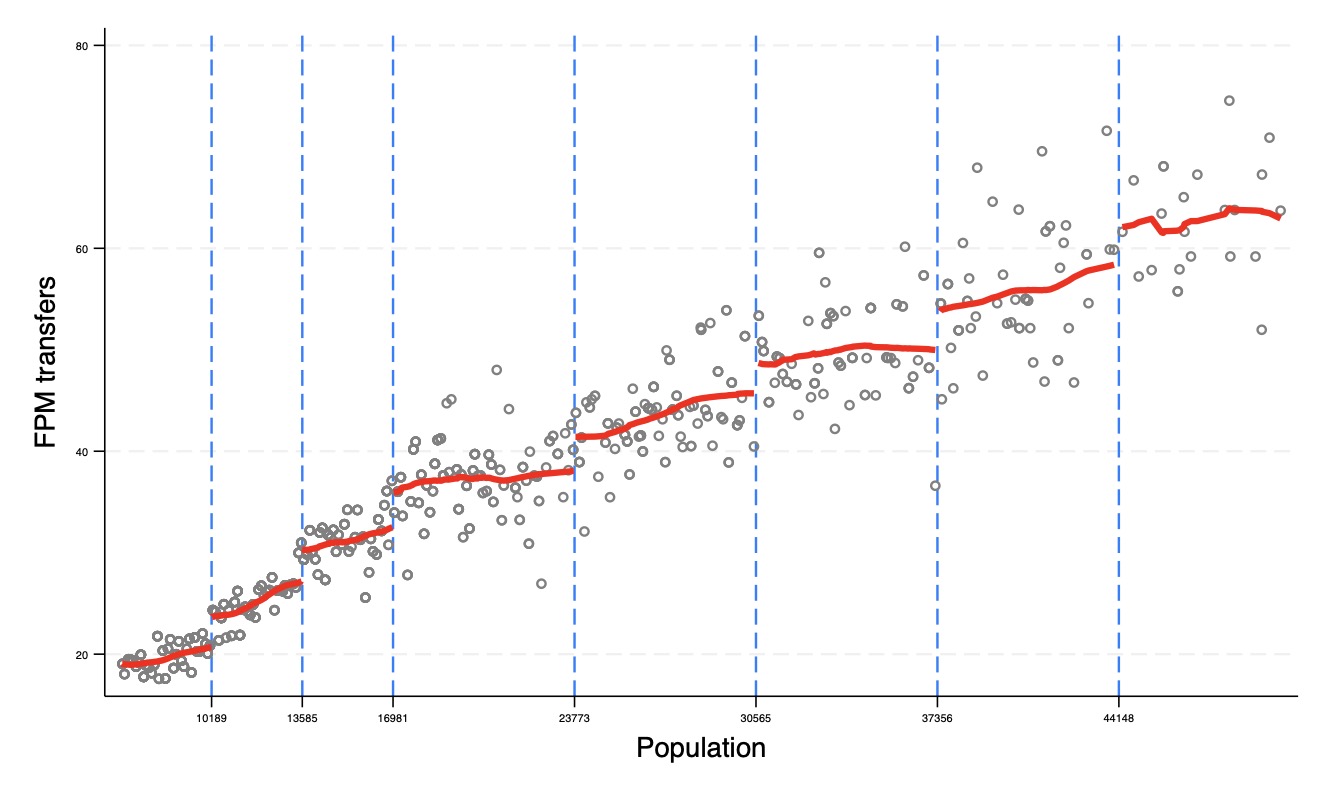}\includegraphics[width=0.5\linewidth]{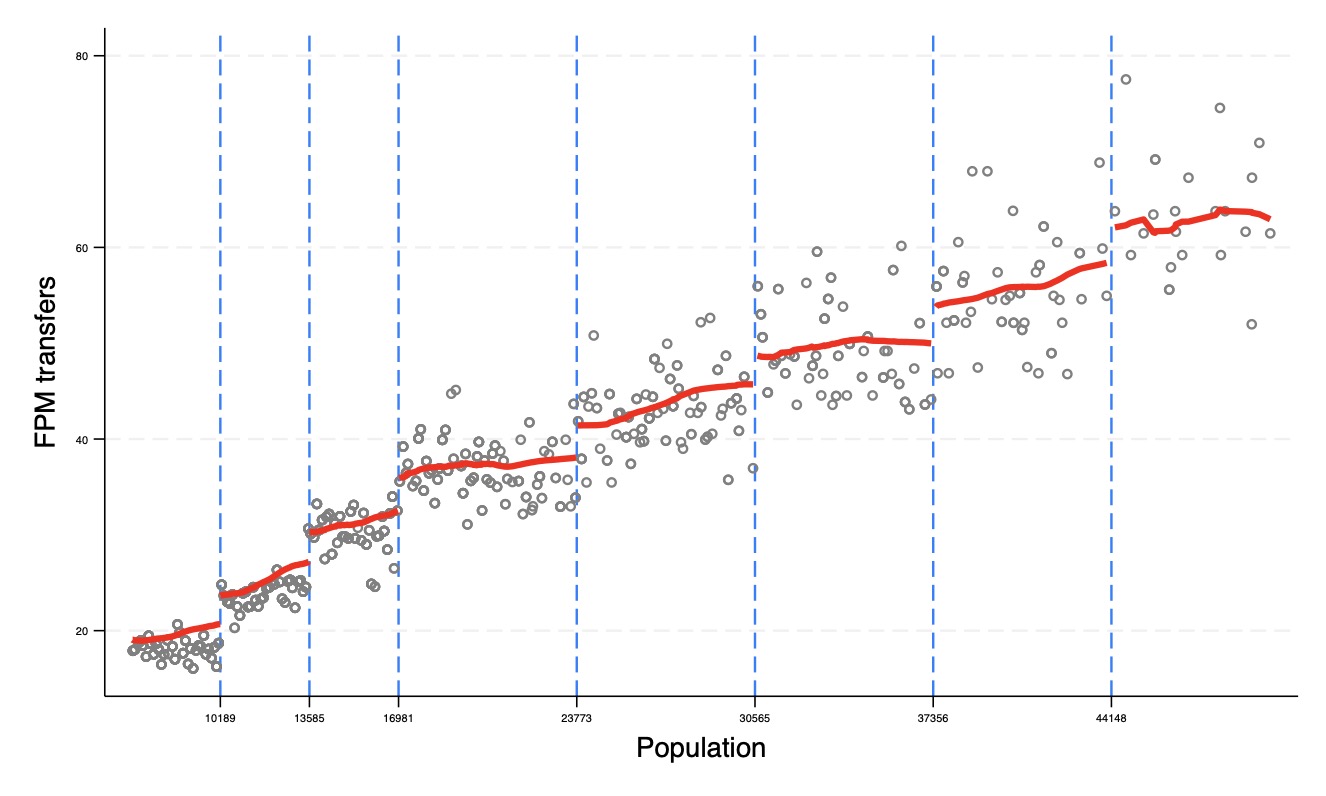}
    \caption{Actual(left) and Theoretical(right) 
Transfers}
    \label{fig:enter-label}
    \label{tab:figure1}
    \label{tab:the figures}
    \textit{Source: BNPT\_small}
\end{figure}

In our model,  $\tau$ denotes the actual transfers, $\hat{\tau}$ denotes the theoretical transfers. \autoref{tab:Table1} shows the FPM population threshold and the associated FPM coefficients, as well as descriptive statistics, by population intervals, on the actual and theoretical FPM transfers. \autoref{tab:figure1} depicts the actual (left panel) and theoretical (right panel) FPM transfers against the population estimates. In each figure, FPM transfers are averaged over cells of 100 inhabitants, plus the smoothed average of transfers (solid line) calculated separately in each interval from one threshold to the next. It is clear that both figures show "sharp" discontinuities at the FPM thresholds. At the same time, theoretical transfers show some within-bracket variability because of the different shares received by each state, and this variability increases with population size. \\

Overall, the visual evidence provided by the figure justifies the necessity for the regression discontinuity design which we will focus on later. The differences between the actual and theoretical transfers suggest the presence of non-compliance or manipulation, vindicating the use of fuzzy RD.\\

\begin{table}[ht]
\centering
\caption{Outcome Measures}
\resizebox{\textwidth}{!}{%
\begin{tabular}{lccccccr}
 & & &\\
\toprule
\thead{Population} & \thead{Broad\\Corruption} & \thead{Narrow\\Corruption} & \thead{Broad\\Fraction\\Amount} & \thead{Narrow\\Fraction\\Amount} & \thead{College} & \thead{Years of\\ schooling} & \thead{Incumbent\\reelection} \\
\midrule
Below 10,189 & 0.79 & 0.37 & 5.69 & 2.17 & 0.40 & 11.52 & 0.31 \\
10,189--13,584 & 0.79 & 0.50 & 5.59 & 1.96 & 0.39 & 11.69 & 0.23 \\
13,585--16,980 & 0.77 & 0.44 & 4.12 & 1.63 & 0.42 & 11.75 & 0.29 \\
16,981--23,772 & 0.82 & 0.55 & 5.79 & 2.68 & 0.43 & 11.70 & 0.26 \\
23,773--30,564 & 0.76 & 0.49 & 5.89 & 2.16 & 0.50 & 12.31 & 0.25 \\
30,565--37,356 & 0.74 & 0.43 & 5.47 & 2.02  & 0.55 & 12.56 & 0.34 \\
37,357--44,148 & 0.77 & 0.39 & 5.20 & 1.96  & 0.50 & 12.54 & 0.38 \\
44,149--50,940 & 0.74 & 0.52 & 2.15 & 1.00  & 0.59 & 12.89 & 0.26 \\
Total & 0.78 & 0.46 & 5.34 & 2.09 &  0.44 & 11.86 & 0.28 \\
\bottomrule
\end{tabular}
}
\endcenter
\textit{Source: BNPT\_small}
\label{tab:Table2}
\end{table}

\clearpage

\section{Measurement Framework}
We will test the FPM discontinuities and the randomness to evaluate the theoretical hypotheses in \textit{the study} following a fuzzy RD and DD setting, then we will briefly touch on the electoral punishment of corruption.
\subsection{Instrumental Variables and Regression-Discontinuity Designs}
As discussed, the empirical framework suggests a clear link between the variation in population size and the treatment assignment, which creates an ideal setup for fuzzy RD design. \\

Namely, FPM Transfers are allocated to municipalities based on population thresholds, with municipalities above each threshold receiving higher theoretical transfers than those below. However, the actual transfers received do not always perfectly align with the theoretical transfers determined by the allocation rule. \\

In a sharp RD design, treatment assignment would be a deterministic function of the population size. In contrast, the fuzzy design accommodates the fact that treatment assignment depends on the \textit{population size }(running variable) in a stochastic manner.This fuzzy nature is evident in \autoref{tab:figure1}, which reveals cases of misassignment around the cutoffs, where some municipalities near a threshold receive transfers inconsistent with their theoretical entitlement.Despite that, the probability of receiving higher or lower transfers(propensity score) still exhibits discontinuities at the population thresholds, it allow us to identify the causal effect of transfers on outcomes of interest, including corruption levels and political indicators. \\

Based on \textit{the study}, we use a potential outcome notation to capture both the outcome of interest ($y_i$), including corruption, politicians’ education, or reelection, and actual transfers depend on theoretical transfers plus other stochastic elements. $y_i(\hat{\tau})$ and $\tau_i(\hat{\tau})$ are the potential values of the outcome and of actual transfers, expressed as a function of theoretical transfers.\\

First stage:
\begin{equation}
\centering
\tau_i = g(P_i) + \alpha_{\tau} \hat{\tau} + \delta_t + \gamma_p + \mu_i
\end{equation}\\

Reduced Form:
\begin{equation}
\centering
y_i = g(P_i) + \alpha_y \hat{\tau}_i + \delta_t + \gamma_p + \eta_i 
\end{equation}\\

$g(P_i)$: high-order polynomial in $P_i$
$\delta_t$: time fixed effects\\
$\gamma_p$: state fixed effects\\
$\mu_i, \eta_i $: error terms clustered at the municipality level\\
$\alpha_{\tau}$: intention-to-treat (IIT) effect of theoretical transfers on actual transfers\\
$\alpha_y$: IIT effect of theoretical transfers on the outcome\\

As we all know, “fuzzy RD is IV”(Angrist and Pischke, 2009).
 Fuzzy RD allows us to estimate the local average treatment effect (LATE). The LATE represents the causal effect of transfers for complier municipalities, those whose treatment status is influenced by the assignment rule. With the instrument we could recreate a randomized experiment's conditions.To evaluate the validity of this IV strategy, we need to assess whether the following key assumptions are satisfied:\\
\begin{enumerate}
    \item Relevance: The instrument (theoretical transfers) must be strongly correlated with the endogenous variable (actual transfers). The result of first-stage regressions presented in (Table3) show a highly significant and positive relationship between theoretical and actual transfers. The F-statistics for the excluded instrument are well above the conventional threshold of 10, indicating a strong first stage and mitigating concerns about weak instrument bias.

    \item Exclusion restriction: The instrument should affect the outcome variables (i.e., corruption, political selection) only through its impact on the endogenous variable (actual transfers). In other words, there should be no direct effect of theoretical transfers on the outcomes, and no omitted variables that are correlated with both the instrument and the outcomes. The FPM allocation rule, which is based on population thresholds, is unlikely to directly influence corruption or political selection, as it is determined by a formula and not by political or economic factors. However, there may be concerns about potential violations of the exclusion restriction if, under exceptional circumstances, the population thresholds coincide with other policy changes that could affect the outcomes. 

    \item Monotonicity: The instrument should affect the endogenous variable in the same direction for all units (municipalities). In the context of the paper, this means that an increase in theoretical transfers should not lead to a decrease in actual transfers for any municipality. It seems plausible given the institutional context and the strong positive correlation between theoretical and actual transfers.

    \item Independence: The instrument should be as good as randomly assigned, meaning that it should not be correlated with potential outcomes or unobserved factors that could influence the outcomes. In the RD design, this assumption is satisfied if municipalities just above and below the population thresholds are similar in all respects except for the amount of transfers they receive. 

    \item In \textit{the original study}, comprehensive tests are implemented, the result suggests that the running variable of our fuzzy RD does not show any evidence of manipulation, so that we can safely use it as a (local) source of exogenous variation.\\
\end{enumerate}

Under the continuity assumption discussed in \textit{the study}, we could write the second stage to identify the causal effect of FPM transfers on the outcome:\\

\begin{equation}
\centering
Y_i = g(P_i) + \beta_{Y_i} T_i + \delta_t + \gamma_p + \epsilon_i
\end{equation}\\

$\epsilon_i $: error terms clustered at the municipality level\\

The causal effects we identify by (3) are local in a twofold meaning. First, because of the RD setup, they only refer to municipalities around the thresholds. Second, because of the IV setup, they only refer to compliers: municipalities that received larger transfers because of the FPM revenue-sharing mechanism. The external validity of our exercise is enhanced by the presence of multiple thresholds. Yet, the identification on compliers leaves aside a subpopulation that might be of interest on its own: the always-takers, that is, municipalities receiving larger transfers irrespective of their position above or below the cutoffs.\\

Note that we control the two-time-period by including time dummies in all specifications and clustering the standard errors at the municipality level, so we could ignore time subscripts in the specification in the RD design\\

We will also delve a little bit deeper into the population estimate ($P_i$) term. Recent research by Imbens et al.(2017) highlighted the potential pitfalls of High-order polynomials, i.e. they can lead to noisy estimates, sensitivity to the degree of the polynomial, or poor coverage of confidence intervals. However, under these potential concerns regarding high-order polynomials, our results turn out to be quite robust, so it's not a big issue. At the same time, alternative methods, including Local regression could induce over-complex issue, so these models are not covered in the discussion.\\

\subsection{Difference in Difference}
To complement the regression discontinuity (RD) analysis, we propose conducting a difference-in-differences (DD) estimation using the panel structure of the data. The DD approach compares the changes in outcomes between municipalities that receive additional transfers (treatment group) and those that do not (control group), before and after the treatment occurs.\\

The DD model specification is as follows:
\begin{equation}
\centering
Y_{i,t} = \alpha + \beta_1 Treat_i + \beta_2 Post_t + \delta_{DD} (Treat_i \times Post_t) + \epsilon_{i,t}
\end{equation}
where:
\begin{itemize}
\item $Y_{i,t}$ is the outcome variable for municipality $i$ at time $t$
\item $Treat_i$ is a dummy variable indicating whether municipality $i$ belongs to the treatment group (i.e., receives additional transfers)
    $=$ 1 if a municipality's actual FPM transfer($\tau_{i,2001}$) exceeds its theoretical entitlement ($\hat{\tau_{i,2001}}$) in the first mayoral term; otherwise $=$ 0
\item $Post_t$ is a dummy variable indicating the post-treatment period, with 0 representing the first mayoral term, and 1 marking the second.
\item $(Treat_i \times Post_t)$ is the interaction term between the treatment and post-treatment dummies
\item $\delta_{DD}$ is the coefficient of interest, capturing the causal effect of the treatment on the outcome
    $\delta_{DD}$=($Y_{treated,2005} - Y_{treated,2001}$)-($Y_{utreated,2005} -Y_{untreated,2001}$)
\item $\alpha$ is a constant term
\item $\epsilon_{i,t}$ is the error term
\end{itemize}

For $Treat_i$, we control for city-specific effect by controlling for for city fixed effects $\gamma_i$, and time fixed effects $\phi_t$:
\begin{equation}
    Y_{i,t} = \alpha + \delta_{\text{DD}}(\text{Treat}_i \times \text{Post}_t) + \phi_t + \gamma_i + \epsilon_{i,t}
\end{equation}

Lastly, we replace the interaction term $(Treat_i \times Post_t)$ by a continuous variable \(\text{Windfall}_{i,t}\), measuring the fraction of municipalities receiving excess political windfalls(FPM):
\begin{equation}
    Y_{i,t} = \alpha + \delta_{\text{DD}} \text{Windfall}_{i,t} + \phi_t + \gamma_i + \epsilon_{i,t}
\end{equation}
Where:
\begin{itemize}
    \item \(\text{Windfall}_{i,t}\) is defined as \([(\tau_i - \hat{\tau}) \times \text{Post}_t]\), representing the excess of FPM received over the theoretical entitlement at mayoral term.
\end{itemize}

In the context of the model, we define the treatment group as municipalities that experience an increase in transfers above a certain threshold (e.g., the median value) from one electoral term to the next. The control group consists of municipalities that do not receive such an increase. The DD coefficient $\delta_{DD}$ captures the average treatment effect on the treated (ATT).

\section{Estimation Results}
\subsection{Regression Discontinuity}

\textbf{The Effect of Federal Transfers on Political Corruption}\\

We start by evaluating the effect of federal transfers on political corruption. The estimation result of the RD design shows significant effects of fiscal windfalls on the frequency of corruption episodes, which is consistent with the theory.\\

\autoref{tab:Table3(a)} estimates the Reduced-Form Effects, namely equation (1) and (2). While \autoref{tab:Table3(b)} investigate the baseline IV regressions, which is equation (3). \\

We focused on the Narrow Fraction Amount, as it captures the proportion of the audited budget that is associated with serious corruption violations. Therefore we are zeroing in on the most egregious forms of corruption, which are more likely to reflect intentional wrongdoing rather than mere administrative irregularities. This provides a clearer conceptual link between fiscal windfalls and rent-seeking behavior.\\

\begin{figure}[t]
    \centering
    \includegraphics[width=0.9\linewidth]{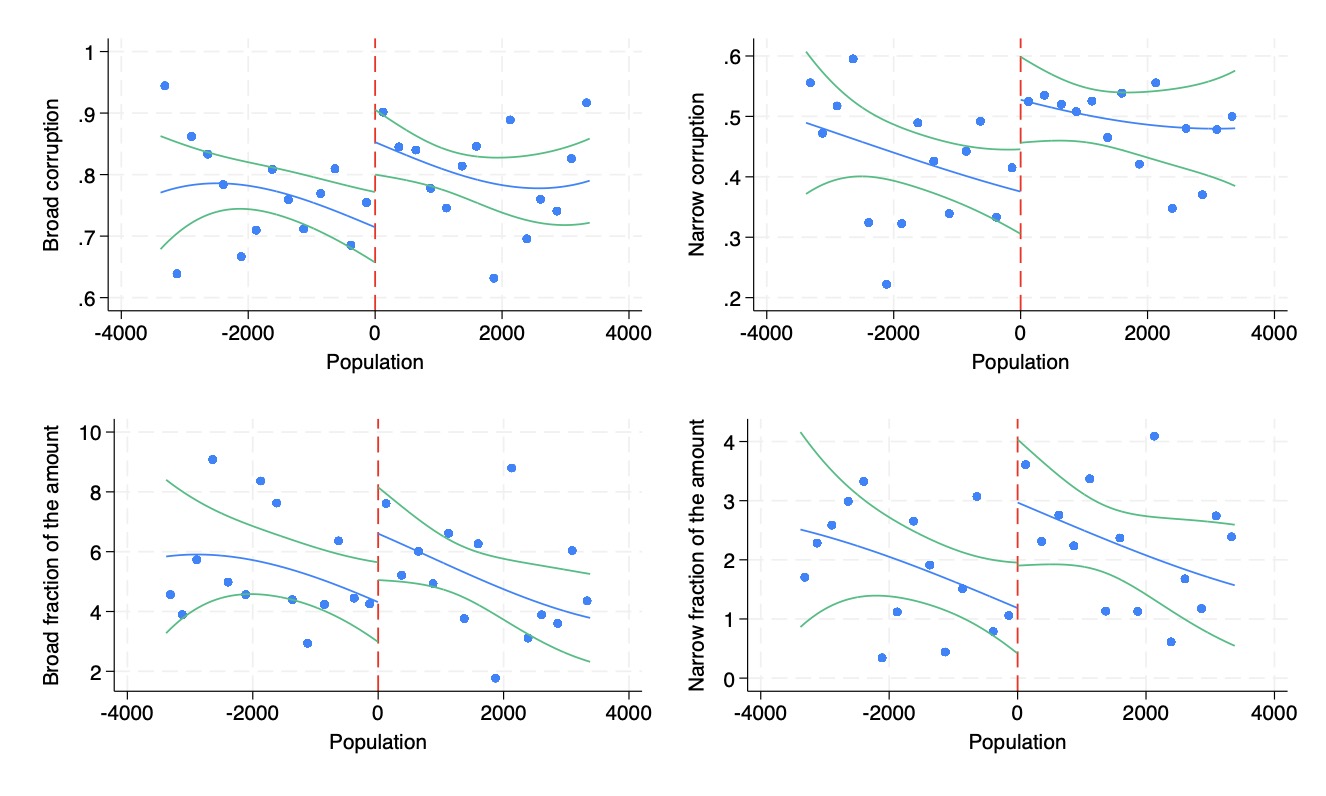}
    \caption{ITT Discontinuities: Corruption Measures}
    \label{fig:enter-label}
    \label{tab:figure2}
    \textit{Source: BNPT\_small}
\end{figure}

In both the reduced-form and IV estimates presented in Table 3, the coefficients for the Narrow Fraction Amount are consistently positive and statistically significant, both for the overall effect and for the pooled thresholds 1-3 and 4-7. This suggests that the relationship between federal transfers and serious corruption is robust. The IV estimates in panel (b) of \autoref{tab:Table3} indicate that a one-standard-deviation increase in actual FPM transfers more than doubles the fraction of the audited budget involved in serious corruption violations. \autoref{tab:figure2} provides graphical proof of relevant and statistically significant discontinuities in the corruption measures induced by the FPM policy. \\

As a conclusion, the estimation results of the RD design confirms that additional revenue triggered by FPM thresholds significantly boost corruption in the municipality. \\

\begin{table}[t]
    \caption{The Effect of Federal Transfers on Political Corruption} 
    \label{tab:effects}
    \begin{subtable}{0.6\linewidth}
      \centering
        \caption{Reduced-Form Effects: FPM Transfers and\\ Corruption Measures}
        \resizebox{\textwidth}{!}{
        \begin{tabular}{lcc}
            \toprule
            & FPM Transfers & Narrow Fraction Amount \\
            \midrule
            Overall effect& 0.637*** & 0.160** \\
            & (0.035) & (0.080) \\
            Thresholds 1-3 & 0.577*** & 0.193*** \\
            & (0.048) & (0.072) \\
            Thresholds 4-7 & 0.651*** & 0.156* \\
             & (0.039) & (0.095) \\
            \bottomrule
            \label{tab:Table3(a)}
        \end{tabular}
        }
    \end{subtable}%
    \begin{subtable}{0.35\linewidth}
      \centering
        \caption{IV Estimates: Corruption\\ Measures}
        \resizebox{\textwidth}{!}{
        \begin{tabular}{lc}
            \toprule
             & Narrow Fraction \\
            \midrule
            Overall effect & 0.245** \\
            & (0.121) \\
            Thresholds 1-3 & 0.322*** \\
            & (0.122) \\
            Thresholds 4-7 & 0.239* \\
            & (0.137) \\
            \bottomrule
            \label{tab:Table3(b)}
        \end{tabular}
        }
    \end{subtable}
    \begin{itemize}
        \item {*** p$<$0.01, ** p$<$0.05, * p$<$0.1}
    \end{itemize}
    \label{tab:Table3}
\end{table}

\textbf{The Effect of Federal Transfers on Political Selection}\\

Following the same logic, we investigate the effect of transfers on the quality of political opponents and on the incumbent’s reelection. Unlike the corruption results, the coefficients here are generally small and statistically insignificant, especially for the overall sample. The reduced-form estimates in \autoref{tab:Table4(a)} show no significant relationship between theoretical transfers and opponents' education or reelection outcomes, except for a marginally significant positive effect on the fraction of opponents with a college degree in the pooled thresholds 4-7. The IV estimates in \autoref{tab:Table4(b)} tell a similar story: there is no robust evidence that actual transfers affect opponents' education or incumbent reelection, although the point estimate for the college variable is again marginally significant for thresholds 4-7. As noted, the sample size in our analysis is smaller than in the original paper, which reduces statistical power and makes it harder to detect significant effects, especially for the heterogeneity analysis across different threshold groups.\\

\begin{table}[t]
    \caption{The Effect of Federal Transfers on Political Corruption} 
    \label{tab:effects}
    \begin{subtable}{0.6\linewidth}
      \centering
        \caption{Reduced-Form Effects: FPM Transfers, \\Opponents’ Education, and Election Outcome}
        \resizebox{\textwidth}{!}{
        \begin{tabular}{lcccc}
            \toprule
            & FPM Transfers & College & Years of Schooling & Incumbent Reelection \\
            \midrule
            Overall effect & 0.637*** & 0.006 & -0.006 & 0.003 \\
            & (0.035) & (0.027) & (0.005) & (0.005) \\
            Thresholds 1-3 & 0.577*** & 0.004 & -0.029 & -0.003 \\
            & (0.048) & (0.039) & (0.006) & (0.006) \\
            Thresholds 4-7 & 0.651*** & 0.007* & 0.016 & 0.003 \\
            & (0.039) & (0.016) & (0.003) & (0.003) \\
            \bottomrule
            \bottomrule
            \label{tab:Table4(a)}
        \end{tabular}
        }
    \end{subtable}%
    \begin{subtable}{0.35\linewidth}
      \centering
        \caption{IV Estimates: Opponents’\\ Education and Election Outcome}
        \resizebox{\textwidth}{!}{
        \begin{tabular}{lccc}
            \toprule
             & College & Years of Schooling & Incumbent Reelection \\
            \midrule
            Overall effect & 0.009 & -0.009 & 0.005 \\
           & (0.006) & (0.042) & (0.008) \\
            Thresholds 1-3& 0.007 & -0.041 & -0.004 \\
            & (0.008) & (0.061) & (0.010) \\
            Thresholds 4-7 & 0.011* & 0.020 & 0.004 \\
            & (0.006) & (0.043) & (0.008) \\
            \bottomrule
            \label{tab:Table4(b)}
        \end{tabular}}
    \end{subtable}
    \begin{itemize}
            \item *** p$<$0.01, ** p$<$0.05, * p$<$0.1
    \end{itemize}
    \label{tab:Table4}
\end{table}

\subsection{Difference in Difference}

\begin{table}[htbp]\centering
\def\sym#1{\ifmmode^{#1}\else\(^{#1}\)\fi}
\caption{Difference-in-Differences Estimation Results}

\begin{tabular}{lc}
 &\\
\hline\hline
                    &\multicolumn{1}{c}{(1)}\\
                    &\multicolumn{1}{c}{broad corruption}\\
\hline
windfall            &      $-$0.671\sym{**} \\
                    &     (0.279)         \\
[1em]
Constant            &       5.228\sym{***}\\
                    &    (0.0486)         \\
\hline
Observations        &        1097         \\
Adjusted \(R^{2}\)  &       0.011         \\
\hline\hline
\multicolumn{2}{l}{\footnotesize Standard errors in parentheses}\\
\multicolumn{2}{l}{\footnotesize Clustered standard errors at the municipality level in parentheses.}\\
\multicolumn{2}{l}{\footnotesize \sym{*} \(p<0.1\), \sym{**} \(p<0.05\), \sym{***} \(p<0.01\)}\\
\end{tabular}
\end{table}

The fixed-effects model exploits within-municipality variation over time to estimate the effect of the continuous treatment variable windfall on the proportion of the audited budget associated with broad corruption violations (fraction\_broad). The estimated coefficient on windfall is -0.6714153, which is statistically significant at the 5\% level (p-value = 0.016). This suggests that, on average, a one-unit increase in the excess transfers received by a municipality (i.e., the difference between actual and theoretical FPM transfers) leads to a 0.6714153 percentage point decrease in the proportion of the audited budget involved in broad corruption violations. The low between R-squared (0.0017) suggests that the treatment variable has limited explanatory power for the cross-sectional variation in corruption levels across municipalities.\\

It is worth noting that the negative coefficient on windfall is somewhat counterintuitive, as it suggests that municipalities receiving more transfers than their theoretical entitlement experience lower levels of corruption. This result could potentially be explained by several factors:

\begin{enumerate}
    \item Measurement error: If the broad corruption measure is subject to significant measurement error, the estimated coefficient may be biased towards zero or even have the wrong sign.

\item Omitted variable bias: Despite the inclusion of municipality     fixed effects, there may be time-varying unobserved factors that are correlated with both the treatment and the outcome variables, leading to biased estimates.
\item Reverse causality: It is possible that municipalities with lower levels of corruption are more likely to receive excess transfers, rather than the other way around. This could happen if higher-level governments allocate additional resources to municipalities that are perceived as less corrupt or more efficient.
\end{enumerate}

\subsection{The Electoral Punishment of Corruption}
In order to investigate whether the electoral punishment for corruption decreases with larger federal transfers, \textit{the study} compare the electoral outcomes of mayors who were audited before the election to those who were audited after the election, and examine how this difference varies with the level of federal transfers.\\

The key specification is a triple difference (DDD) model:
\begin{align*}
E_i = & \ \beta_1 (\hat{T}_i \cdot \text{before}_i \cdot y_i) + \beta_2 (\text{before}_i \cdot y_i) + \beta_3 (\hat{T}_i \cdot \text{before}_i) \\
      & + \beta_4 (\hat{T}_i \cdot y_i) + \alpha_1 \hat{T}_i + \alpha_2 y_i + \alpha_3 \text{before}_i \\
      & + g(P_i) + g(P_i) \cdot \text{before}_i \cdot y_i + g(P_i) \cdot \text{before}_i + g(P_i) \cdot y_i \\
      & + \delta_t + \gamma_p + \psi_i
\end{align*}

Utilising this specification, we could briefly touch on the impact of FPM transfers on the punishment of corruption and we focus on incumbent runs for reelection. As shown in \autoref{tab:tab3}, The key variables of interest are:

\begin{enumerate}
    \item The interaction between the "before" dummy (indicating whether the audit was released before the election) and the corruption measures. This captures the effect of disclosing corruption on the probability of running for reelection.

\item The "super" interaction between the "before" dummy, the corruption measures, and the theoretical FPM transfers. This captures how the effect of disclosing corruption on the probability of running for reelection varies with the level of transfers.
\end{enumerate}
The results in Panel A show that the disclosure of corruption before the election has a negative effect on the probability of running for reelection, as indicated by the negative coefficients on the interaction terms "Before × corruption" across all corruption measures. The logic behind this analysis is that if larger transfers reduce the electoral punishment for corruption, incumbent mayors in municipalities with higher transfers should be more likely to run for reelection even when their corrupt activities are exposed before the election. The positive coefficients on the triple interaction terms, while not always significant, provide some support for this hypothesis.

\begin{table}[t]
\centering
\caption{Impact of FPM Transfers on the Punishment of Corruptionn}
\begin{tabular}{lcccc}
 & & &\\
\toprule
 Incumbent runs for reelection

 & \thead{Broad \\corruption}& \thead{Narrow \\corruption} & \thead{Broad \\corruption \\Amount}& \thead{Narrow \\corruption \\Amount}\\ \hline
 &  &  &  &  \\
Before × corruption × FPM& -0.009 & 0.166 & 0.007 & 0.015 \\
 & (0.125) & (0.112) & (0.004) & (0.015) \\
Before × corruption& -2.471*** & -1.482** & -0.050* & -0.106 \\
 & (0.747) & (0.588) & (0.029) & (0.093) \\
Before × FPM& 0.122 & -0.011 & 0.083 & 0.082 \\
 & (0.116) & (0.090) & (0.071) & (0.069) \\
 Observations & 789 & 789 & 740 & 740 \\
 R-squared & 0.082 & 0.088 & 0.077 & 0.076 \\ \hline
\multicolumn{5}{l}{ Robust standard errors in parentheses} \\
\multicolumn{5}{l}{ *** p$<$0.01, ** p$<$0.05, * p$<$0.1} \\
\end{tabular}
\label{tab:tab3}
\end{table}

\section{Conclusion}
This paper revisits a "political resource curse" framework proposed by Brollo et al.. So far, we have focused on two fundamental mechanisms in political institutions: the effects of additional resources on political corruption and on the incentives to participate in politics. In light of these theoretical results, we have investigated a specific Brazilian institution that provides an ideal quasi-experimental setting, and we found considerable support for the implications of the theory.

\section{Appendix}

\subsection{The Anti-Corruption Program and Measurements of Corruption}
The Brazilian anti-corruption Program initiated in 2003 randomly selects municipalities to be audited by the Controladoria Geral da União (CGU), an independent body that assesses the use of federal transfers at the local level. The audit reports generated by the CGU form the basis for the study's corruption indicators. These reports contain information on the total amount of federal transfers audited. Because of the random selection process, the program provides a valuable source of information on budget irregularities and corruption episodes in municipal governments. \\

We are mainly concerned about how an exogenous increase in FPM transfers around the population thresholds affects corruption in the use of all sources of municipal revenues. Since FPM transfers account for the largest fraction of municipal revenues(40\%), and 70\% of FPM transfers are unrestricted, we can safely test how benefits grabbed by politicians(corruptions) react to a change in overall budget size $\tau$ (caused by the windfall). According to the theory discussed in \textit{the study}, as FPM transfers increase, municipal governments feel less restrained in pleasing the voters and results in increasing corruption of all kinds, and not just corruption concerning the FPM transfers. We follow the classification of narrows and broad corruption described in \textit{the study}.\\

To proceed, we refer to “BNPT\_small,” consisting of 1,157 observations, as the random sample for which we have information on the corruption measures. \autoref{tab:Table2} reports descriptive statistics on these variables by population thresholds. Broad and Narrow corruption are dummy variables determining the type of corruption. Fraction Amounts measure the amount of corruption. 

\subsection{Measuring the Quality of Politicians}
We measure the Quality of Politicians with reference to education(average features of the pool of candidates in each municipal election included in our sample). As showed in \autoref{tab:Table2}, College is the fraction of opponents holding a college degree. Years of schooling measures the opponent’s average years of schooling. Incumbent reelection is the probability of the incumbent being reappointed. \\

\section{Reference}
Brollo, Fernanda, Tommaso Nannicini, Roberto Perotti, and Guido Tabellini. “The Political Resource Curse.” The American Economic Review 103, no. 5 (2013): 1759–96. \\http://www.jstor.org/stable/42920629.\\

Gelman, Andrew, and Guido Imbens. \textit{“Why High-Order Polynomials Should Not Be Used in Regression Discontinuity Designs.”} Journal of Business \& Economic Statistics 37, no. 3 (2019): 447–56. doi:10.1080/07350015.2017.1366909.\\

Angrist, Joshua D., and Jörn-Steffen Pischke. \textit{Mostly Harmless Econometrics: An Empiricist’s Companion}. Princeton University Press, 2009. https://doi.org/10.2307/j.ctvcm4j72.

\end{document}